\documentclass[%
reprint,
longbibliography,
groupedaddress,
amsmath,amssymb,longbibliography,
aps
prl,
]{revtex4-2}

\usepackage{hyperref}
\usepackage{graphicx} 
\usepackage{dcolumn}
\usepackage{float}

\usepackage{bm}

\newcommand{\bec}[1]{\mbox{\boldmath $ #1$}}

{}
{}
{}
{}
{}
{}
{}
{}
{}
{}
{}
{}

{}
{}
{}

\newcommand{\ovec}[1]{{\mbox{\boldmath $#1$}}}
\newcommand{\bscE}{\ovec{\cal{E}}}

\usepackage{xcolor}

\begin{document}

\title{Neural Differential Equations for the Solar Dynamo}

\author{E. Illarionov}
 \email{egor.illarionov@math.msu.ru}
 \affiliation{Moscow State University, Moscow, Russia}
 \affiliation{Institute of Continuous Media Mechanics, Perm, Russia}

\author{R. Stepanov}
\affiliation{Institute of Continuous Media Mechanics, Perm, Russia}
\affiliation{Perm National Research Polytechnic University, Perm, Russia}

 \author{K. M. Kuzanyan}
 \affiliation{IZMIRAN, Troitsk, Moscow, Russia}
 \affiliation{Institute of Continuous Media Mechanics, Perm, Russia}

\author{V. Kisielius}
\affiliation{Moscow State University, Moscow, Russia}
\affiliation{Institute of Continuous Media Mechanics, Perm, Russia}

\begin{abstract}
Physical models aimed to reproduce basic features of the solar sunspot cycle are typically based on the solar dynamo mechanism.
Usually qualitative arguments are used to define parameters of the model, among which a challenging component is the nonlinear form of quenching of the alpha-effect governing regeneration of the magnetic field.
We propose a novel approach, in which the functional form of the alpha-quenching is represented by a neural network model embedded into neural differential dynamo equations trained on observational data. For demonstration, we consider a low-mode dynamo model and 
find a wide set of alpha-quenching functions and corresponding dynamo numbers that provide an accurate fit to the average profile of the solar cycle data given by sunspot numbers. Within this set, we observe a strong relationship  between the dynamo number and the shape of the alpha-quenching function indicating that additional magnetic field data or constraints are essential to unambiguously infer parameters of the dynamo model.  
In our opinion, the neural differential approach opens a new prospect for data-driven investigation of the closure problem in dynamo theory.
\end{abstract}


\maketitle

\section{Introduction}

A fully resolved simulation of the solar dynamo, capturing all relevant spatial and temporal scales, remains beyond current computational capabilities. Constructing a tractable dynamo model, based on physical mechanisms, consistent with observations, and with predictive power is a challenging task. The mean field theory provides a useful framework for description of the effects of magnetohydrodynamic (MHD) turbulence on generation of the large scale magnetic field due to a turbulent electromotive force \cite{Krause1980,2006PhRvE..73e6311R}. This approach allows one to obtain different types of dynamo models applicable to different astrophysical objects, estimate generation thresholds, and interpret the results of dynamo and MHD turbulence experiments. \cite{2001PhRvL..86.3024G,2007PhRvL..98d4502M,2010PhRvL.105r4502F}.  

Formulation of mean-field dynamo models faces the problem of relation of the mean electromotive force to the mean magnetic field (the closure problem) that becomes much more complicated in a fully nonlinear regime, where the velocity field at small scales is influenced by the large-scale magnetic field. To address this problem, phenomenological relations in algebraic form and quenching of turbulent coefficients are usually proposed, based on qualitative arguments or theoretical derivation  (e.g., \cite{Rogachevskii2000}). But the models may still include a large number of free parameters that require calibration, and the assumptions about the properties of turbulence have to be justified.


Estimation of the model parameters using post-processing of direct numerical simulations of MHD turbulence (e.g., \cite{Pipin2011}) or using simple search (e.g., Monte Carlo simulations, \cite{Newton2013}) 
have been used to adopt the model to reproduce observational data. 
Apparently, the first data-driven attempt was recently carried out in \cite{Bonanno_2025}, where the dynamics was inferred using sparse regression on a set of basis functions. 

Our study will focus on one of the central and most important aspects of mean field dynamo mechanism: the nonlinear quenching of the alpha-effect, which plays a critical role in the saturation of magnetic field growth. The alpha-quenching function will be treated as an unknown to be inferred directly from observational data using the framework of neural differential equations (hereafter, NDE, \cite{NDE}). This approach bypasses phenomenological assumptions and provides a data-driven reconstruction of the nonlinear feedback.


The key idea is to apply NDE to adjust theoretical physical models (given as a system of differential equations) to observational data or, more generally, to optimize some objective. 
Actually, methods for estimation of parameters in ordinary differential equations (ODE) are quite well elaborated (see, e.g., \cite{DICKINSON1976123}, \cite{Cao}, and recent reviews \cite{Sapienza}, \cite{Lettermann}). The key feature of the NDE framework is that when the parameters are unknown functions, they can be represented using neural networks that can capture complex relations, and the trainable parameters of the neural networks are considered as the parameters of the ODE and are estimated accordingly.
An attempt to collect various applications of this approach has been recently proposed in \cite{NDEreview}. The most relevant applications to solar physics are carried out in turbulence simulation (see \cite{Williams, Boral}).

In this Letter we demonstrate the feasibility of the NDE approach using a simple low-mode solar dynamo model that still captures basic features of the cyclic solar magnetic activity. 
We argue that this proof-of-concept work establishes a novel approach to resolving the closure problem in mean-field dynamo theory and illustrates the broader potential of neural differential equations to bridge the gap between physical modeling and observational data 
analysis in complex astrophysical systems.

\section{Neural Differential Dynamo Model}

We develop a neural differential dynamo model based on the eminent Parker dynamo \citep{Parker1955}. Similar models \cite{1991A&A...245..654S,1997SoPh..173....1K} have been utilized to describe the generation of the poloidal $A$ and toroidal $B$ components of the axisymmetric magnetic field in a thin spherical shell.
We further reduce the problem by decomposing the solution into a Fourier series of meridional modes and keeping only the largest spatial mode. Thus, we arrive to the simplest possible low-mode $\alpha\Omega$ dynamo model \cite{Newton2013}, whose dimensionless ordinary differential equations are
\begin{eqnarray}
\begin{gathered}
  \dot{B}(t) = -i D A(t) - B(t), \\
  \dot{A}(t) = \alpha(B(t)) B(t) - A(t),
\label{eq:dynamo}
\end{gathered}
\end{eqnarray}
where  $D=G \alpha_0 R_\odot^4/\beta^2$ is a dynamo number, and function $\alpha(B)$ is the $\alpha$-effect. 
The equations are non-dimensionalised using the turbulent magnetic diffusivity $\beta$, the characteristic differential rotation $G$, the solar radius $R_\odot$, the characteristic $\alpha$-effect $\alpha_0$,  $B_0$ and $B_0 \alpha_0 R_\odot^2/\beta$ for $B(t)$ and $A(t)$, respectively. 
%

For the linear case with constant $\alpha(B)=1$, the model possesses exponentially growing oscillatory solutions if the dynamo number satisfies the condition $D > 2$.
When nonlinear quenching is introduced using the typical algebraic function $\alpha(B) = 1/(1 + ReB^2)$ (see, e.g., \cite{Charbonneau2020}), the model solution saturates to stationary oscillations, reflecting the nonlinear back reaction of the magnetic field to hydrodynamic helicity.

To combine the dynamo model \eqref{eq:dynamo} with the NDE framework,
the set of initial values $A_0 = A(t_0)$, $B_0 = B(t_0)$, and dynamo number $D$ are considered as unknown parameters to be optimized. The unknown function $\alpha(B)$ is parametrized using a fully-connected neural network model. Let $\theta$ denote the trainable parameters of this neural network, then $\alpha=\alpha(B
\,|
\,\theta)$. The full set of the dynamo model parameters is now $A_0, B_0, D, \theta$.

\begin{figure}[h]
\centering
\includegraphics[width=0.48\textwidth]{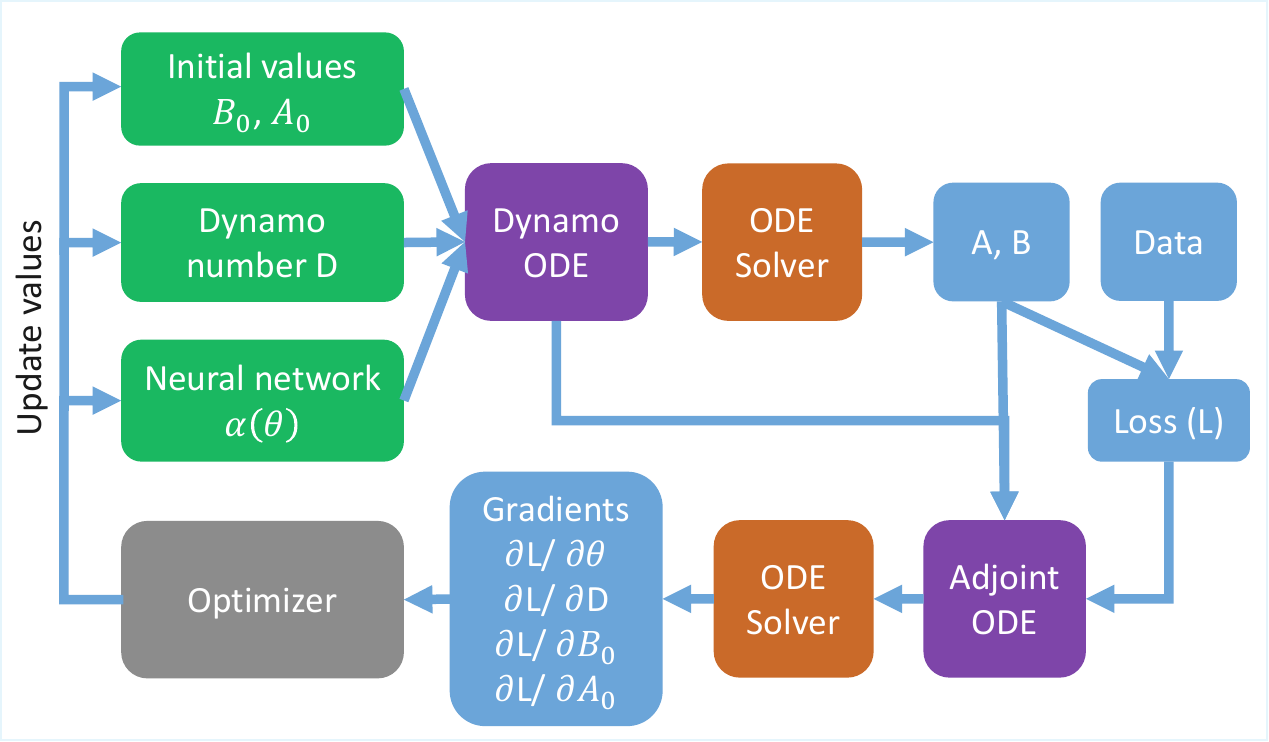}
\caption{Scheme of the neural differential framework. Green boxes contain parameters to be optimized. Current values of these parameters are passed to the dynamo model and the ODE solver yields the numerical solutions $A$, $B$. Numerical solutions are compared with observational data and the loss function is computed. Based on the dynamo ODE, solutions $A$, $B$, and the loss function, the adjoint ODE is constructed (see Appendix A for an example). Integration of the adjoint ODE using the ODE solver yields the gradients of the loss function with respect to the model parameters. Optimizer uses the gradients to update the model parameters.}
\label{fig:nde_scheme}
\end{figure}

Now consider a time interval $[0, T]$ and assume that we have observational data $d_0, d_1, ..., d_n$ corresponding to time moments $0=t_0 < t_1 < t_2 < ... < t_n=T$. The generic form of the loss function is $L(A(t_0), B(t_0), d_0, A(t_1), B(t_1), d_1, ..., A(t_n), B(t_n), d_n)$. As an example, if observational data are associated with the energy of the real part of the magnetic field $B$, then the loss function can be formulated as the sum of squared errors:
\begin{equation}
    L(ReB) = \sum\limits_{i=0}^n\left(ReB(t_i)^2 - d_i\right)^2.
\label{eq:loss}
\end{equation}
The value of the loss function depends on the parameters $A_0, B_0, D, \theta$. Using the adjoint method to compute gradients of the loss function with respect to these parameters and applying gradient descent method (specifically, we will use Adam optimization algorithm \cite{Adam}), we arrive to a set of values for these parameters that provide a (local) minima for the loss function. In particular, we obtain the optimum dynamo number $D$ and the form of the function~$\alpha$. The whole optimization pipeline is shown in Figure~\ref{fig:nde_scheme}.

\section{Synthetic experiments}

We validate the NDE approach on a set of synthetic examples, in which we first obtain a numerical solution of the dynamo equations \eqref{eq:dynamo} with some prescribed dynamo parameters (referred to as ground truth), then, given the ODE solution only, reconstruct the dynamo parameters using NDE, and compare the reconstructed parameters with ground-truth ones. 

To be specific, let us generate the solution of the dynamo equations \eqref{eq:dynamo} on a time interval $[0, 40]$ with $A(0)=1$, $B(0)=1$, $D=5$, and $\alpha(B) = 1 /(1 + ReB^2)$ using a fixed-size time step $\Delta t=0.1$ (these settings yield six magnetic field cycles). 

Now consider the neural differential dynamo model initialized with "random" dynamo parameters and initial values. Of course, an arbitrary initial guess for dynamo parameters could lead to an exploding solution, so we pick up a random value from a reasonable range of the parameters. In particular, we sample
$ReA_0, ImA_0, ReB_0, ImB_0$ uniformly in $[-1, 1]$, $D$ uniformly in $[3,10]$. 
Regarding $\alpha$, we assume that it depends on $ReB$ only, is symmetric, non-negative, and is normalized so that  $\alpha(0)=1$. 

An obvious choice is to represent the function $\alpha$ as a feed-forward neural network model. However, the non-negativity and normalization conditions require a special configuration of the model. The non-negativity condition is usually achieved by applying an appropriate activation function to the output layer of a neural network. In contrast, forcing the model to give a pre-described value for a particular input value is less trivial. This can be achieved by adding term $(\alpha(0)-1)^2$ to the loss function but requires balancing this term with other terms in the loss function and does not ensure a perfect match. Another option is to redefine the model so that $\tilde{\alpha}(\cdot)=\alpha(\cdot)/\alpha(0)$ but this complicates backpropagation and makes the training of $\tilde{\alpha}$ less stable. Instead, we multiply the output of the model by its input (this ensures that the output is zero when the input is zero) and apply an activation function that is non-negative and equals unity when the input is zero. In particular, the exponential activation function is a  valid option. This results in the model of the form $\alpha(B)=\exp(ReB^2g(ReB^2))$, where $g(\cdot)$ is an unconstrained neural network.

We select a simple feed-forward neural network $g(\cdot)$ consisting of fully connected layers with the number of neurons 1, 4, 4, 1 in successive layers. This configuration is one of the simplest capable of capturing the standard smooth functions within which we expect to find the actual alpha function. In total, it contains 24 weighting and 9 bias parameters. The hidden layers are followed by activation function
 $ELU(x)=x{\rm I}\{x>0\} + (\exp(x)-1){\rm I}\{x\le0\}$, where ${\rm I}\{\cdot\}$ is the indicator function (see \cite{DUBEY202292} for a review of common activation functions). The output layer has no activation function. More complex architectures should be considered if one expects complex variations in function $\alpha$.

A non-necessary but practically useful step is to pre-train the neural network $\alpha(B)$ so it matches some simple analytical function, for which the solution of the dynamo equations does not explode. Specifically, we pre-train $\alpha(B)$ so that it matches $exp(-ReB^2)$ in the interval $ReB^2\in[0, 10]$. Realization of this step is trivial, and we do not discuss it in details.

\begin{figure}[h]
\centering
\includegraphics[width=0.48\textwidth]{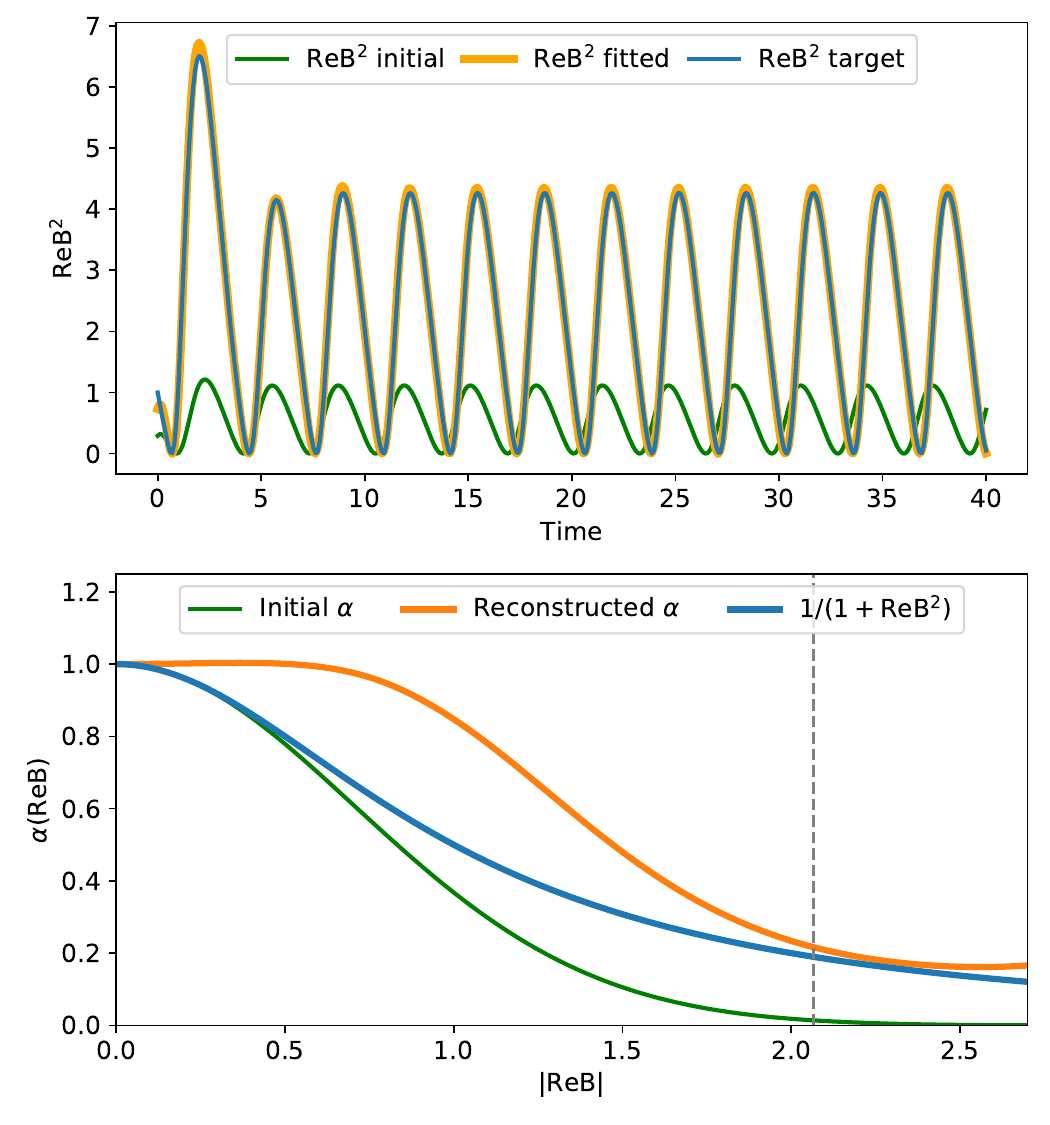}
\caption{One realization of the reconstruction of the dynamo parameters when only $ReB^2$ is fitted. Upper panel shows a comparison of the initial state of the model (green line), fitted model (orange line) and target (blue line) values of $ReB^2$. Bottom panel shows the initial function $\alpha$ (green line), reconstructed (orange line) and ground-truth (blue line) function~$\alpha$. Dashed gray line shows the maximum value of $|ReB|$ after stabilization of the dynamo model.}
\label{fig:4panels}
\end{figure}

We start by demonstrating that the NDE framework is capable of finding ODE parameters that provide an accurate fit to the data when only a rough initial guess is given for these parameters. In Figure~\ref{fig:4panels}, we reconstruct dynamo parameters fitting only $ReB^2$. Upper panel shows that we can obtain an accurate phase-amplitude match starting from a rough initial guess. Bottom panel shows that this fit can be achieved with $\alpha$ that is significantly different from the ground-truth function except for the values near the extreme value of $|ReB|$ (indicated by gray dashed line). 
 In this case, the reconstructed dynamo number $D$ is 3.71 (initial "random" $D$ is 3.36), reconstructed initial values are $A_0=0.96+0.45i$, $B_0=0.85-1.40i$.

Given that we can reconstruct dynamo parameters that provide the required fit but are different from the ground-truth ones, the next step is to determine how much the dynamo parameters may vary, and how this variability changes when we fit the parameters using more information on the magnetic field data, not just $ReB^2$.
To address it, we consider three ways to reconstruct the dynamo parameters: (i) fitting only $ReB^2$, 
(ii) fitting both $ReB^2$ and $ReA$, and (iii) fitting the full set of $ReB, ImB, ReA, ImA$. In each case, we repeat the optimization process 100 times starting from various independent initial values for the dynamo parameters. One technical note is that when optimizing several fields, which may have different scales, we found it useful to normalize them to the same scale (this step is typical in machine learning).

\begin{figure}[h]
\centering
\includegraphics[width=0.48\textwidth]{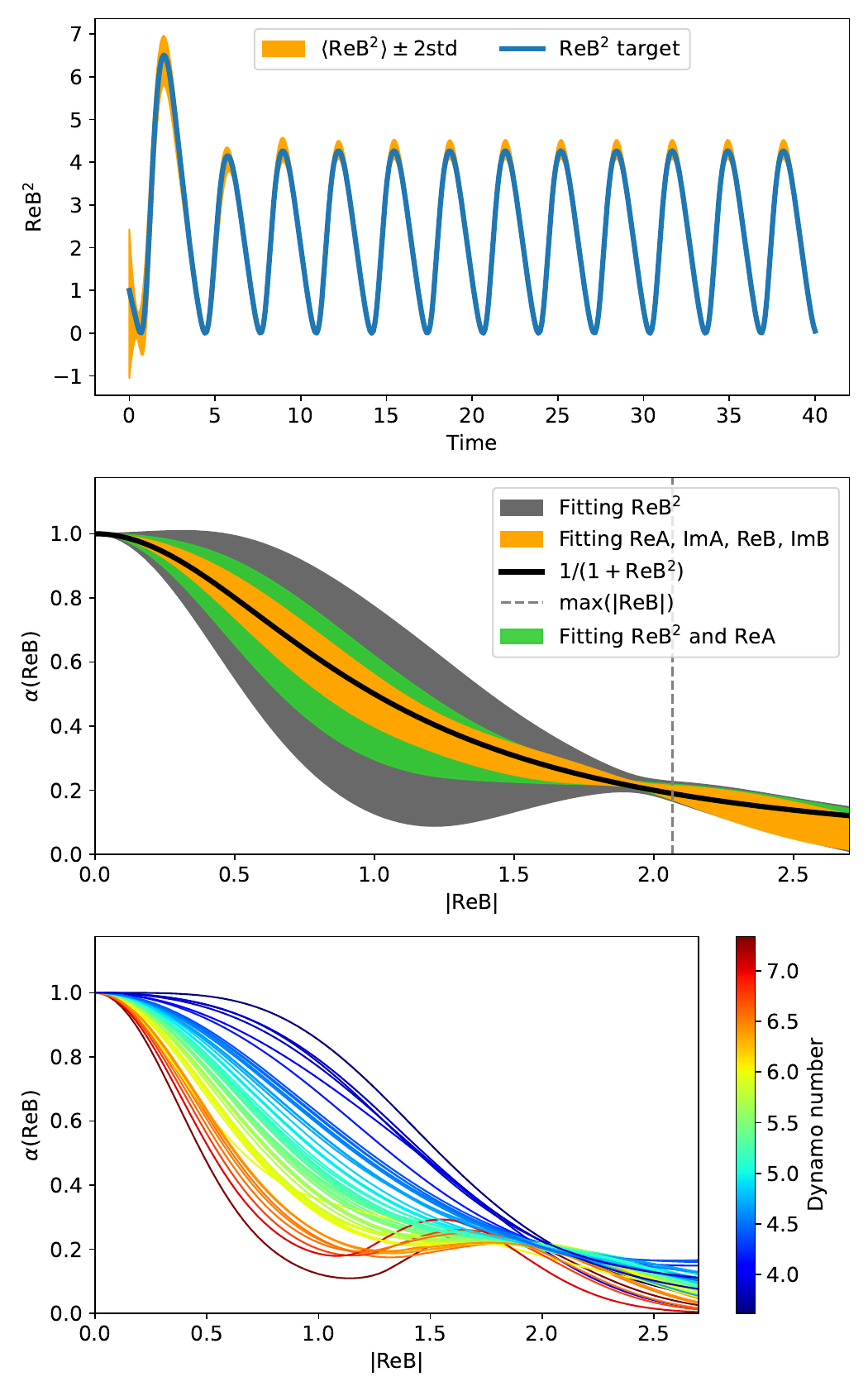}
\caption{Upper panel: variability (two-sigma interval) of the fitted models (orange area) and the ground-truth values of $ReB^2$. Middle panel: variability  (two-sigma intervals) of reconstructed functions $\alpha$ when (i) only $ReB^2$ is fitted (gray area), (ii) both $ReB^2$ and $ReA$ are fitted (green area), and (iii) full set $ReA$, $ImA$, $ReB$, $ImB$ is fitted. Black line shows the ground-truth function $\alpha(B)=1/(1+ReB^2)$.
Bottom panel: individual reconstructed functions $\alpha$ when only $ReB^2$ is fitted, colored according to the value of the corresponding dynamo number.}
\label{fig:alpha_cases}
\end{figure}

Upper panel in Figure~\ref{fig:alpha_cases} shows that all considered realizations provide accurate fit of the data (we show only $ReB^2$, which is common in all three cases). The only notable deviations are in the beginning of the integration interval, and there are minor deviates near the maxima of the cycles. Middle panel shows that among these realizations, larger variability of functions $\alpha(B)$ arises when we fit $ReB^2$ only, and the variability becomes lower when $ReB^2$ is complemented with additional magnetic field data. Importantly, we observe that when a more complete set of magnetic field data is provided to the model, the reconstructed functions $\alpha$ converge to the ground-truth one.
Also in each case the width of the uncertainty interval  becomes small when $|ReB|$ approaches its highest values.

The corresponding variability of the dynamo number $D$ is given in Table~\ref{tab:dyn}. Similar to $\alpha$, the variability reduces when a more complete set of magnetic field data is provided to the model, and the estimates converge to the ground-truth value.

Bottom panel in Figure~\ref{fig:alpha_cases} shows the relationship between the set of reconstructed functions $\alpha$ (when only $ReB^2$ is fitted) and corresponding dynamo numbers. We observe that larger dynamo values correspond to a faster decay of function $\alpha$; when the dynamo number is large enough, function $\alpha$ becomes non-monotonic. 
The latter result clearly demonstrates the non-uniqueness of the solution to the inverse nonlinear problem.


\begin{table}[]
    \caption{Variability of the dynamo number $D$ (mean value and standard deviation) over 100 independent optimization runs.}
    \begin{ruledtabular}
    \begin{tabular}{lcc}
         Case & Mean $D$ & 
         $\sigma(D)$ \\
         \hline
         Fitting $ReB^2$ & 5.31 & 0.83 \\
         Fitting $ReB^2$ and $ReA$ & 5.20 & 0.41 \\
         Fitting all $ReA$, $ImA$, $ReB$, $ImB$ & 5.10 & 0.26 \\
         \hline
         Ground-truth & 5 &
    \end{tabular}
    \end{ruledtabular}
    \label{tab:dyn}
\end{table} 

\section{Mean solar cycle}

\begin{figure}[h]
\centering
\includegraphics[width=0.48\textwidth]{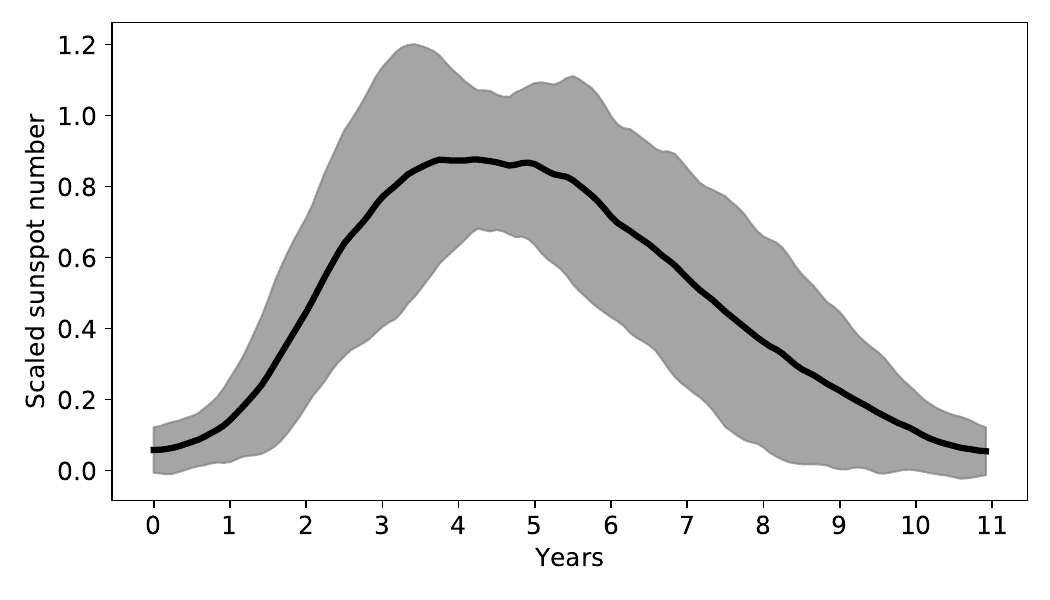}
\caption{Averaged profile of the last 24 solar cycles (black line) and region of two standard deviation from the average (gray area).}
\label{fig:mean_cycle}
\end{figure}

The behavior of observed solar activity is significantly more complex than the model solution. Solar cycles are quasi-periodic, meaning that they are characterized by irregular changes in intensity and duration. Realistic chaotic dynamics is beyond the scope of the low-mode dynamo model \eqref{eq:dynamo}.
However, we can try to capture the profile of an average cycle from the available observational data using the monthly sunspot numbers (SN) from the World Data Center SILSO, Royal Observatory of Belgium \citep{SILSO}. 
We applied a two-year Gaussian filter to unambiguously determine the minima and maxima of solar activity, identifying 24 complete cycles. Next, the duration of all cycles was scaled to 11 years and the intensity was scaled to one unit. Figure~\ref{fig:mean_cycle} shows profile of the averaged solar cycle and sample variation. One can see faster growth and slower decay, which is a robust feature of solar activity \cite{Usoskin}. 
The physical causes that produce this regularity are not fully understood. 
In particular, deviations of the shape of solar and stellar cycles from the harmonic one can be related to nonlinear effects (see, e.g., \cite{Baliunas_ea2006}) and fluctuations in dynamo models (see, e.g., \cite{Dmitrieva_ea2000, Karak2011, Pipin2011ApJ}).
In the framework of our model, we expect to obtain agreement with asymmetry of the solar cycle profile using a complex dynamo number and an appropriate choice of the quenching function $\alpha(B)$.

Then we repeated the mean cycle profile 12 times and scaled to the time interval $[0, 40]$ to make the setup compatible with those used for synthetic experiments.

We associate the obtained mean cycle profile with $ReB^2$ (similar to, e.g., \cite{Passos2008}), and conduct optimization process aimed to minimize the squared difference between $ReB^2$ and mean cycles 100 times sampling initial values in the same way as we did in the synthetic experiments. In contrast to the synthetic case, we allow $D$ to be a complex number that helps to adjust the cycle period.

\begin{figure}[H]
\centering
\includegraphics[width=0.48\textwidth]{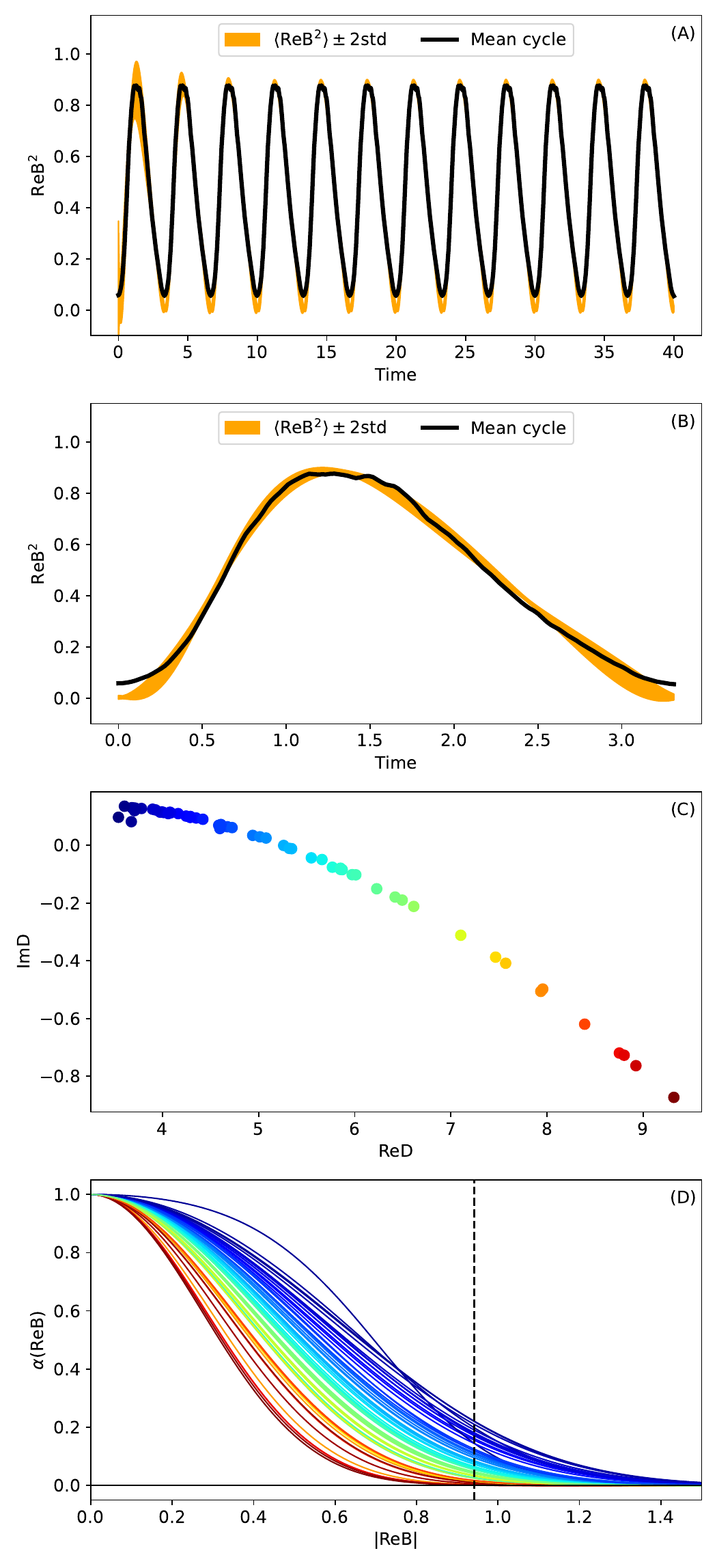}
\caption{Panel~A: variability of the fitted models (orange area) compared to mean cycle profiles (black line). Panel~B: last cycle cropped from the upper panel. Panels~C and D: variability of the fitted dynamo numbers shown in the complex plane (panel C) and variability of the corresponding functions $\alpha$ (panel D). Similar colors in panels C and D indicate corresponding pairs of parameters of the fitted models. Dashed gray line in panel D shows the maximal value of cyclic variation of $|ReB|$.
}
\label{fig:4panels_cycle}
\end{figure}

Figure~\ref{fig:4panels_cycle} shows that NDE is able to fit the period and amplitude of the cycles as well as the asymmetric form of the cycle.
Similar to synthetic cases, we observe a large variability of functions $\alpha(B)$ and dynamo numbers $D$ that provide such accurate fit. In contrast to the synthetic cases, the variability is not reduced near the extreme value of $|ReB|$. We assume that it is due to the fact that D is complex, which allows for larger variability in $\alpha(B)$. We observe that in the complex plane, the fitted values of $D$ tend to lie on a smooth curve. Moving $D$ along this curve, the corresponding function $\alpha(B)$ also changes smoothly, and larger $ReD$ lead to a faster decrease of $\alpha(B)$ similar to the synthetic case.

\section{Discussion and Conclusions}

In this work, we demonstrate the feasibility of using a neural network to adapt a low-mode dynamo model to observational data. Our implementation successfully reconstructs the form of nonlinear quenching of the $\alpha$-effect, one of the most challenging mechanisms in dynamo theory to address both theoretically and numerically.

Importantly, the training procedure remains effective even with reduced datasets, indicating that the essential features of the nonlinear feedback can be recovered without requiring extensive or high-resolution data.

Our approach also enables a systematic assessment of the sensitivity of dynamo solutions to the functional form of $\alpha$-quenching. This is particularly valuable for validating closure models in mean-field dynamo theory, where phenomenological prescriptions are often adopted with limited justification. The ability to infer quenching functions directly from data thus opens a new route for bridging theory and observations in solar and astrophysical dynamos.

The resulting dynamo solutions (see Figure~\ref{fig:4panels_cycle}) reproduce basic features of typical observational data. For instance, the growth phase of the solar cycle is generally shorter than the decay phase, a property that our model accurately captures. At the same time, we observe a large variability in the reconstructed dynamo parameters and a strong relationship between them. We assume that this variability can be reduced either by using additional magnetic field data (as we demonstrated in synthetic case), or by introducing additional constraints in the dynamo model.

Of course, applying NDE framework still requires careful tuning to achieve robust results. For example, the gradient descent process can occasionally converge to parameter sets for which the ODE solution diverges, such as when the learning rate is too high or the initial guess is poorly chosen. The architecture of the neural network represents a natural compromise between the variability of representable functions and the difficulty of optimization. Training the neural network model requires a solution of the adjoint ODE, which in the case of our model took about four times more computational time than numerically solving the forward system of the dynamo model equations. Since each training iteration consists of one forward and one backward pass, it becomes equivalent to about five forward integrations in total. Of course, this scaling may depend on the dynamo models and on the performance of the particular framework implementing the NDE.


There is significant potential for further development of this method. Natural extensions for the case of solar dynamo include increasing the number of spatial modes in the dynamo model, introducing explicit latitudinal dependence, and incorporating nonlinear magnetic contributions to the $\alpha$-effect. Another important direction is training on spatially resolved observational or simulation data, which would test the scalability of the neural differential equation framework in more realistic settings. Finally, one can introduce time dependence into the dynamo model parameters and fit the time-dependent parameters using the adjoint method. If the obtained parameters change more slowly than the solar cycle period, they can be used to extrapolate (predict) the solar cycle.

\section*{Data availability}

The data and source code that support the findings of this article are available in GitHub repository \cite{github}.

\begin{acknowledgments}
The research is supported by RSF grant 21-72-20067. The authors acknowledge the reviewers for valuable comments and suggestions and the Lomonosov-2 supercomputer center at MSU for providing computational resources.
\end{acknowledgments}

\appendix

\section{Neural differential equations}

Neural differential equations (NDE) is a framework that 
combines neural networks and differentiable ODE solvers, allowing
optimization of functional terms in ODEs. To understand this union, it is useful to consider neural networks as a function parametrized with a set of parameters $\theta$, also called trainable weights (with a certain parameterization, this function can approximate continuous functions with any desired accuracy, see Cybenko's universal approximation theorem \cite{Cybenko}). Embedding the parametric function in the ODE, we obtain an ODE of the form
\begin{eqnarray}
\begin{gathered}
    \dot{y} = f(y, t, \theta), \\
    y(t_0) = y_0.
\label{eq:ode}
\end{gathered}
\end{eqnarray}

Let $y(t)$ be a numerical solution of this ODE, $L(y)$ be an objective function (loss function in machine learning) constructed from ODE solution, and we want to optimize this objective function (typically, minimize) with respect to the parameters $\theta$ using gradient descent methods.




Direct computation of gradient of the loss function $L$ with respect to $\theta$ requires differentiating the solution of the ODE, which is impossible treating the ODE solver as a black box. However, one can avoid it using the adjoint method grounded in the Pontryagin’s maximum principle  \cite{Pontryagin1962}. The idea for the simplest case, where $L(y)=L(y(t_1))$, is that the gradient
${{\rm d} L}/{{\rm d}\theta}$ is shown to be obtained as the integral
\begin{equation}
    \frac{{\rm d} L}{{\rm d}\theta} = \int\limits_{t_0}^{t_1}\lambda(t)^T\frac{\partial f}{\partial \theta}{\rm d}t,
\end{equation}
where $\lambda(t)$ is called the adjoint state and is obtained as the solution of the reverse-time ODE
\begin{eqnarray}
\begin{gathered}
    \dot{\lambda} = -\frac{\partial f}{\partial y}\lambda, \\
    \lambda(t_1) = \frac{\partial L}{\partial y(t_1)}.
\label{eq:adjoint_ode}
\end{gathered}
\end{eqnarray}

Note that to integrate the adjoint ODE, $y(t)$ should be known on the whole integration interval.  The derivation of the adjoint method is given in \cite{NDE} and the implementation 
is given in the repository \url{https://github.com/rtqichen/torchdiffeq}.
Considering the specific ODE augmentation, one can obtain gradients with respect to initial condition $y_0$ as well as to $t_0$ and $t_1$, the start and end times of the integration interval (see also \cite{NDE}).

To summarize, the adjoint method implemented in differentiable ODE
solvers allows optimizing the parameters of an ODE according to a given objective function. When these parameters are trainable weights of a neural network model, this is called neural differential equations. 


\bibliography{ref}

\end{document}